\documentclass[%
 reprint,
 amsmath,amssymb,
 aps,
]{revtex4-2}

\usepackage{graphicx}
\usepackage{dcolumn}
\usepackage{bm}
\usepackage{amsfonts}
\usepackage{bbold}
\usepackage{pgfplots}
\pgfplotsset{compat=1.14}
\usepackage[font=small,labelfont=bf,justification=justified,format=plain]{caption}
\usepackage[utf8]{inputenc}
\usepackage[right]{lineno}
\usepackage[justification=justified,format=plain]{subcaption}

\newcommand{\w}{\omega}
\newcommand{\e}{\epsilon}
\newcommand{\g}{\gamma}
\newcommand{\la}{\lambda}

\newcommand{\bp}{\langle\Psi_+|}
\newcommand{\kp}{|\Psi_+\rangle}
\newcommand{\bmm}{\langle\Psi_-|}
\newcommand{\km}{|\Psi_-\rangle}
\begin{document}

\preprint{APS/123-QED}

\title{Kubo conductivity for anisotropic tilted Dirac semimetals and its application to  8-Pmmn borophene:\\ The role of different frequency, temperature and scattering limits}

\author{Sa\'ul A. Herrera}
\author{Gerardo G. Naumis}%
 \email{naumis@fisica.unam.mx}
\affiliation{%
Depto. de Sistemas Complejos, Instituto de F\'isica, Universidad Nacional Aut\'onoma de M\'exico\\
Apdo. Postal 20-364, 01000, CDMX, M\'exico.
}%


\date{\today}

\begin{abstract}
The electronic and optical conductivities for  anisotropic tilted Dirac semimetals are calculated using the Kubo formula. As in  graphene, it is shown that the  minimal conductivity is sensitive to the order in which the temperature, frequency and scattering limits are taken.  Both intraband and interband scattering are found to be direction dependent. In  the high frequency and low temperature limit, the conductivities do not depend on frequency and are  weighted by the anisotropy in such a way that the geometrical mean $\sqrt{\sigma_{xx}\sigma_{yy}}$ of the conductivity  is the same as in graphene. This results from the fact that in the zero temperature limit, interband transitions are not affected by the tilt in the dispersion, a result that is physically interpreted as a global tilting of the allowed transitions. Such result is verified by an independent and direct calculation of the absorption coefficient using the Fermi golden rule. However, as temperature is raised, an interesting minimum is observed in the interband scattering, interpreted here as a result of the interplay between the tilt and the chemical potential increasing with temperature.
\begin{description}
\item[Keywords]
Dirac semimetal, 8-Pmmn borophene, Kubo conductivity.
\end{description}
\end{abstract}

\maketitle


\section{\label{sec:level1}Introduction}
In the last years, Dirac and Weyl semimetals have attracted intense research interest \cite{Vozmediano2015,Naumis1, Soluyanov2015,Armitage,Yan,Vozmediano2016,Charlier2018,Mele2017,Carbotte2016,Carbotte2018} after the discovery of the one-atom-thick (2D) carbon allotrope, graphene \cite{Novoselov666,Castro1}, showing great promise for applications in the next generation of nanoelectronics \cite{Castro1,Naumis1,Naumis2009,Naumis2007,Sarma}.
After the discovery of graphene, much work has been directed towards searching for new 2D materials which can host massless Dirac fermions \cite{Zhou,Geng,Ye,AndradeNaumis2019-1,AndradeNaumis2019-2}.
In more recent times, 2D crystalline boron allotropes, known as borophenes, have attracted intense research interest due to their chemical and structural complexity \cite{Zhou,Penev2012,Penev2016,Mannix,Ni,Li,NaumisChampo}. Remarkably, a two dimensional phase of boron with space group \textit{Pmmn} was theoretically predicted  to host massless Dirac fermions \cite{Zhou,Lopez}.

8-Pmmn borophene is a 2D boron allotrope known to host massless Dirac fermions with an anisotropic, tilted energy dispersion \cite{Zhou,Lopez,Zabolotskiy} which is found to lead to direction-dependent electronic behavior \cite{Sadhukhan,Zhang2018,Verma,SierraNaumis2019}, a situation akin to strained graphene \cite{Roman2014,Roman2015,Oliva_Leyva_2015,Oliva2016,Naumis1}. Its lattice is formed by a sublattice of ``inner'' atoms and a sublattice of ``ridge'' atoms\cite{Lopez}. A possible origin of the tilt on 8-Pmmn borophene's energy dispersion could be the structure of the inner sublattice, which resembles that of quinoid-type strained graphene, known to present a tilted energy dispersion \cite{Goerbig,Trescher}. However, there seems to be a lack of consensus regarding whether it is the inner sublattice which is mainly responsible for the formation of the Dirac cones or rather both sublattices contribute equally \cite{Zhou, Lopez, Zabolotskiy}.
\begin{figure}[b]
\includegraphics[width=0.45\textwidth]{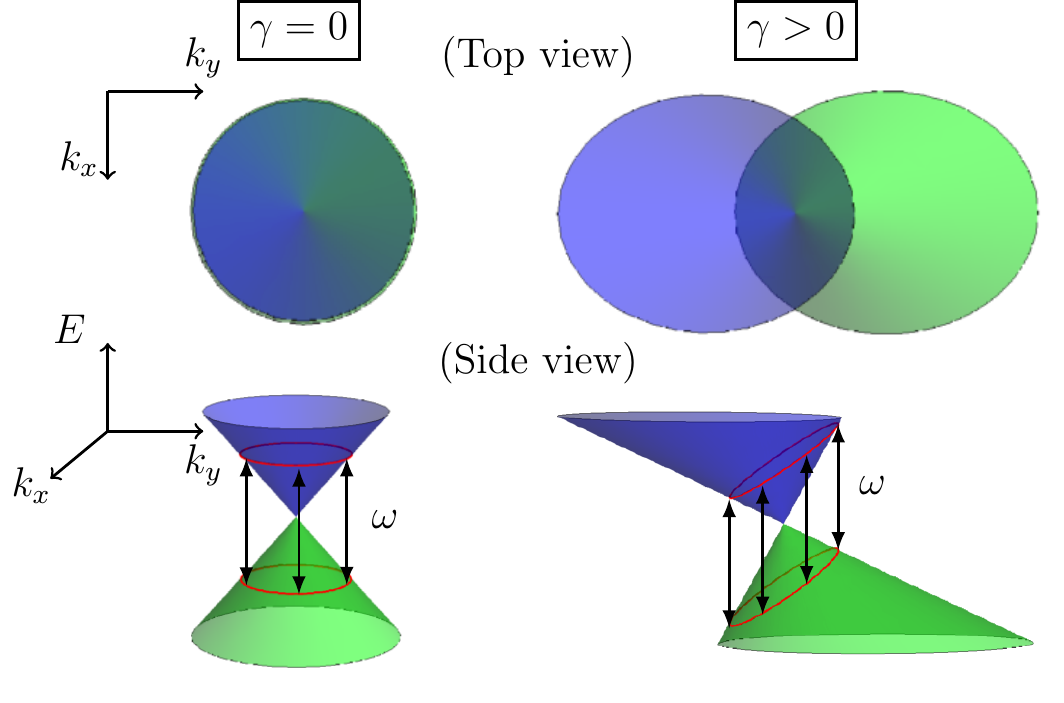}
\caption{\label{Fig:DispersionCone} Comparison of a graphene-like dispersion with a tilted energy dispersion. Direct interband transitions on a tilted dispersion cone for a given constant frequency $\omega$ of the electric field arise from a non-isoenergetic curve, in contrast to transitions in a non-tilted dispersion.}
\end{figure}

In many 2D systems as in 8-Pmmn borophene,  around one of the Dirac points, the low-energy excitations are described by an effective anisotropic tilted Dirac Hamiltonian of the form \cite{Zabolotskiy}
\begin{equation}\label{Eq:Hamiltonian}
    H=v_x\sigma_x k_x+v_y\sigma_y k_y+v_t\sigma_0 k_y.
\end{equation}
Here {\it all energies are in units of} $\hbar$. The other valley is studied through changing the sign of two velocities \cite{Zabolotskiy}.

As seen in Fig. \ref{Fig:DispersionCone}, the last term in Eq. (\ref{Eq:Hamiltonian}) with nonzero $v_t$ produces a tilting of the dispersion cone in the $y$ direction. Also, as $v_x\neq v_y$ and $v_t\neq0$, the cone's constant energy contours are found to be elliptical rather than circular as in the particular case of graphene, for which $v_x=v_y$ and $v_t=0$. These distortions of the dispersion cones of 8-Pmmn borophene are found to produce direction dependent terms and scaling factors in the conductivity, both absent in graphene.

Although there are now some works in which the zero-temperature conductivity in anisotropic tilted Dirac Hamiltonians is calculated through the Kubo formula \cite{Suzumura,Suzumura2,Verma,Goerbig2019}, not so much attention has been drawn to the role of a non-zero temperature, or to the fact that different results depend upon different physical and mathematical limits. This is well known in graphene's optical \cite{Mak,Ziegler,Naumis1} and electronic properties \cite{Mak,Ziegler,Zhang2005,Naumis1}.
For example, in the low-frequency limit, graphene's optical conductivity depends upon the sample through self-doping, scattering, temperature and ripple effects \cite{Mak,Naumis1}.

The aim of this work is precisely to further investigate all the previous effects in the conductivity when applied to anisotropic tilted Dirac Hamiltonians, and specially to 8-Pmmn borophene. In particular, we are interested in the physical understanding of anisotropy and tilting effects on the intra and inter band scattering, which as we will show here, are far from trivial. 

The layout is the following: In section \ref{Sec. sigma calculation} we calculate the conductivity: in subsection \ref{Subsec.Sigmayy} the calculation is done along the direction parallel to the tilt axis, in subsection \ref{Subsec. Sigmaxx} we calculate the conductivity in the direction perpendicular to the tilt axis and contrast with the results of subsection \ref{Subsec.Sigmayy}. Then, in section \ref{Sec.MinimalConductivity} we analyze the different frequency, temperature and scattering limits, and the minimal conductivities obtained thereof, followed by the discussion of the results in section \ref{Sec.Discussion}. Finally, the conclusions are given in subsection \ref{Subsec. Conclusions}.

\section{\label{Sec. sigma calculation}Conductivity for the anisotropic tilted Dirac Hamiltonian}

Let us calculate the conductivity for the Hamiltonian given by Eq. (\ref{Eq:Hamiltonian}). As represented in Fig. \ref{Fig:DispersionCone}, the eigenvalues of this Hamiltonian are given by  $E_{\pm}(\boldsymbol{k})=v_t k_y\pm \sqrt{v_x^2k_x^2+v_y^2k_y^2}$ with eigenvectors 
\begin{equation}\label{Eq.Eigenvectors}
    |\Psi_{+}(\boldsymbol{k})\rangle=\frac{1}{\sqrt{2}}
    \left( \begin{array}{ccc}
1   \\
+e^{i\theta}  \end{array} \right),\;\;\;
|\Psi_{-}(\boldsymbol{k})\rangle=\frac{1}{\sqrt{2}}
    \left( \begin{array}{ccc}
1   \\
-e^{i\theta}  \end{array} \right)
\end{equation}
with $\theta$ defined by $\cos\theta=k_x/|k|$ and $\sin\theta=k_y/|k|$. 
Our aim is to obtain an expression for the real part of the diagonal conductivity from the Kubo formula \cite{Madelung,Ziegler},
\begin{eqnarray}
\label{Eq:sigmadef}
\sigma_{\nu\nu}=\pi \frac{e^2}{\hslash}&&\int \text{Tr}\{[H,r_\nu]\delta(H-\epsilon-\omega)\nonumber\\
&&\times[H,r_\nu]\delta(H-\epsilon)\}\frac{f_\beta(\epsilon+\omega)-f_\beta(\epsilon)}{\omega}d\epsilon
\end{eqnarray}
where $r_{\nu}$ is the position coordinate in the $\nu=x$ or $\nu=y$ direction, $f_\beta(\epsilon)=1/[1+exp(\beta\epsilon)]$ is the Fermi distribution with $\beta=1/k_BT$, $T$ being the temperature and $\delta(x)$ is the Dirac delta function of $x$. In the limit $\beta \rightarrow \infty$, after changing $\epsilon\rightarrow\epsilon -\omega/2$, we obtain, 
\begin{eqnarray}\label{Eq.sigmadefzeroT}
    \sigma_{\nu\nu}=-\pi \frac{e^2}{\hslash}\frac{1}{\omega}&&\int_{-\omega/2}^{\omega/2} \text{Tr}\{[H,r_\nu]\delta(H-\omega/2-\epsilon)\nonumber\\ &&\times[H,r_\nu]\delta(H+\omega/2-\epsilon)\}d\epsilon 
    \end{eqnarray}
    
 Since the current operator is given by  $j_{\nu}=-ie[H,r_{\nu}]$, the trace in Eq. (\ref{Eq.sigmadefzeroT}) can be expressed as,
\begin{eqnarray}\label{Eq.TraceDef}
    T(\epsilon) =\int \text{Tr}_2 \biggl[\frac{\partial H}{\partial k_\nu}&&\delta(H-\omega/2-\epsilon) \nonumber \\&&  \times\frac{\partial H}{\partial k_\nu}\delta(H+\omega/2-\epsilon)\biggr]\frac{d^2k}{(2\pi)^2},
\end{eqnarray}
where $\text{Tr}_2$ is the trace taken over the  pseudospin degree of freedom. In order to investigate the effects introduced by the anisotropy on the electronic properties of this Hamiltonian, we calculate the components of the real conductivity in the direction of the tilt ($\sigma_{yy}$), and in the direction perpendicular to the tilt ($\sigma_{xx}$). These two cases are considered in the following subsections. Notice that for simplicity, here we start considering only one valley and spin. To find the total conductivity, on needs to take into account the corresponding factors, as we do in Section \ref{Sec.Discussion}.

\subsection{\label{Subsec.Sigmayy} Conductivity in the direction parallel to the tilt}

To obtain $\sigma_{yy}$, we start by writing the current operator in the $y$ direction,
\begin{equation}\label{Eq.CurrentOpDef}
    j_y=-ie[H,y]\;\;\rightarrow\;\; e\frac{\partial H}{\partial k_y}=e
    \left( \begin{array}{ccc}
v_t & -iv_y \\
iv_y & v_t   \end{array} \right)
\end{equation}
In order to evaluate the trace in Eq. (\ref{Eq:sigmadef}), we rewrite $j_y$ in the $\{|\Psi_+\rangle,|\Psi_-\rangle\}$ basis. It reads,

\begin{equation}\label{Eq.CurrentOp}
j_y=\mathcal{U}\Big(e\frac{\partial H}{\partial k_y}\Big)\mathcal{U}^\dagger=
 e v_y\left( \begin{array}{ccc}
\gamma +\sin\theta & i\cos\theta \\
-i\cos\theta & \gamma-\sin\theta \end{array} \right).
\end{equation}

To simplify the calculation, we next propose a transformation which defines scaled momentums in such a way that the anisotropy due to $v_x$ and $v_y$ can be eliminated. Thus we define $\xi_x=v_x k_x$, $\xi_y=v_y k_y$, $\xi=\sqrt{\xi^{2}_x+\xi^{2}_y}$ and $\gamma=v_t/v_y$.
Using these new variables, the Hamiltonian is written as, 
\begin{equation}\label{Eq:HamiltonianMatrix}
    H =  \left( \begin{array}{ccc}
\gamma\xi_{y} & \xi_x-i\xi_y  \\
\xi_x+i\xi_y & \gamma\xi_{y}   \end{array} \right),
\end{equation}

Here $\gamma$ serves as a measure of the tilt in the dispersion cone, which in the present Hamiltonian occurs in the $y$ direction. The energy dispersion is now,
\begin{equation}\label{Eq.EnergyDispersion}
    E_{\pm}(\boldsymbol{k})=\gamma\xi_y\pm\xi= (\gamma \sin \theta \pm  1)\xi.
\end{equation}
where $\xi_y=\xi\sin\theta$ is in polar coordinates.
Next we calculate $T(\epsilon)$ using Eq. (\ref{Eq.TraceDef}), and as we show in  Appendix \ref{Appendix}, for $\nu=y$ we obtain that
\begin{widetext}
 \begin{eqnarray}\label{Eq:T_yy}
    T(&&\epsilon)=\int \biggl\{ \frac{(v_t\xi+v_y \xi_y)^2}{\xi^2}\delta(\xi+\gamma\xi_y-\bar{\epsilon}_{+})\delta(\xi+\gamma\xi_y+\bar{\epsilon}_{-}) 
    +\frac{v_y^2\xi_x^2}{\xi^2}\delta(\xi-\gamma\xi_y+\bar{\epsilon}_{+})\delta(\xi+\gamma\xi_y+\bar{\epsilon}_{-}) \nonumber \\
    &&+\frac{v_y^2\xi_x^2}{\xi^2}\delta(\xi+\gamma\xi_y-\bar{\epsilon}_{+})\delta(\xi-\gamma\xi_y-\bar{\epsilon}_{-}) + \frac{(v_t\xi-v_y \xi_y)^2}{\xi^2}\delta(\xi-\gamma\xi_y+\bar{\epsilon}_{+})\delta(\xi-\gamma\xi_y-\bar{\epsilon}_{-})\biggr\}
    \frac{d\xi_x d\xi_y}{v_x v_y(2\pi)^2},
 \end{eqnarray}
 \end{widetext}
 where $ \bar{\epsilon}_{\pm}=\omega/2\pm \epsilon$. This equation reduces to the case of graphene \cite{Ziegler} for $\gamma=0$ and $v_x=v_y$. However, as a consequence of the tilting ($\gamma\neq0$), $T(\epsilon)$ is no longer symmetric under $\epsilon \rightarrow -\epsilon$, rather, it is symmetric under $\epsilon \rightarrow -\epsilon$, $\xi_y\rightarrow-\xi_y$.

We denote the terms in the above integral as $T(\epsilon)=T_{+}^{tra}(\epsilon)+T_{+}^{ter}(\epsilon)+T_{-}^{ter}(\epsilon)+T_{-}^{tra}(\epsilon)$. The superscript $tra$ is used to denote intraband contributions while  the superscript $ter$ denotes interband contributions. The first term of the previous equation, in polar coordinates reads

\begin{eqnarray}\label{Eq.T1}
        T_+^{tra}(\epsilon)=
        \int_0^{2\pi}&&\int_0^\lambda \frac{v_y}{v_x}(\gamma+ \sin\theta)^2\delta(\xi(1+\gamma\sin\theta)-\bar{\epsilon}_{+}) \nonumber \\   
        &&\times\delta(\xi(1+\gamma\sin\theta)+\bar{\epsilon}_{-})\frac{\xi d\xi d\theta}{(2\pi)^2}
\end{eqnarray}

after using $\xi_y=\xi\sin\theta$, $\gamma=v_t/v_y$ and $d\xi_x d\xi_y \rightarrow \xi d\xi d\theta$, and having introduced $\lambda$ as a high energy cutoff \cite{Ziegler}. Similar definitions are used for the other three terms $T_{+}^{ter}(\epsilon),T_{-}^{ter}(\epsilon), T_{-}^{tra}(\epsilon)$. We introduce a scattering rate $\eta$ by considering soft Dirac delta functions $\delta_\eta(x)$
\begin{equation}\label{Eq.DeltaDef}
    \delta(x)\approx\delta_\eta(x)=\lim_{\eta\rightarrow 0}\frac{1}{\pi}\frac{\eta}{x^2+\eta^2}.
\end{equation}
As we will integrate over $\xi$ before $\theta$, we define $\gamma_+=1+\gamma\sin\theta$ and then express the first Dirac delta in Eq. (\ref{Eq.T1}) as
\begin{equation}\label{Eq.DeltaFactor}
    \delta_\eta(\gamma_+\xi-\bar{\epsilon}_{+})=\frac{1}{\gamma_+}\delta_{\eta_+}\Big(\xi-\frac{\bar{\epsilon}_{+}}{\gamma_+}\Big)
\end{equation}
with $\eta_+=\eta/\gamma_+$. For these and further equations to remain well defined we will assume that $0 \le \gamma<1$, which in the three dimensional case defines a type-I Weyl semimetal \cite{GoerbigtypeII,Soluyanov2015}.

We will further make the assumption that $\gamma$ does not take values too close to unity so $\eta_\pm \rightarrow 0$ remains valid. This means that $\delta_{\eta_\pm}(x)$ stays as a good approximation to a (soft) Dirac delta function so we can consider  $\delta_{\eta_\pm}(x)\approx \delta_\eta(x)$, by taking $\gamma\eta \rightarrow 0$.

After having defined  $\gamma_-=1-\gamma\sin\theta$,  $\eta_-=\eta/\gamma_-$, and expressing the delta functions as in Eq. (\ref{Eq.DeltaFactor}), the four terms in the trace in Eq. (\ref{Eq:T_yy}) can be divided into two intraband contributions,
\begin{equation}\label{Eq.T tra}
    T_{\pm}^{tra}(\epsilon)=\int_0^{2\pi}\int_0^\lambda
    g_{\pm}(\theta)\delta_{\eta} \Big(\xi \mp\frac{\bar{\epsilon}_{+}}{\gamma_{\pm}}\Big)\delta_{\eta} \Big(\xi \pm \frac{\bar{\epsilon}_{-}}{\gamma_{\pm}}\Big)\frac{ \xi d\xi d\theta}{(2\pi)^2},
\end{equation}
and two interband contributions,
\begin{equation}\label{Eq.T ter}
    T_{\pm}^{ter}(\epsilon)=\int_0^{2\pi}\int_0^\lambda  l(\theta)\delta_{\eta}\Big(\xi \pm\frac{\bar{\epsilon}_{+}}{\gamma_{\mp}}\Big)\delta_{\eta}\Big(\xi \pm \frac{\bar{\epsilon}_{-}}{\gamma_{\pm}}\Big)\frac{ \xi d\xi d\theta}{(2\pi)^2},
\end{equation}

where,
\begin{equation}
    g_{\pm}(\theta)=\frac{v_y}{v_x}(\gamma\pm\sin\theta)^2 \frac{1}{\gamma_{\pm}^2}
\end{equation}
and,
\begin{equation}
    l(\theta)=\frac{v_y}{v_x}\frac{\cos^2\theta}{\gamma_+\gamma_-}
\end{equation}

The radial integrals in eqs. (\ref{Eq.T tra}-\ref{Eq.T ter}) involving the product of soft Dirac delta functions can be expressed as
\begin{eqnarray}\label{Eq.DoubleDeltaInt}
        \int_0^\lambda && \delta_\eta(\xi-a)\delta_\eta(\xi-b)\xi d\xi\sim(a+b)\delta_\eta\left(a-b\right)\frac{1}{4}[\Theta(\lambda-a) \nonumber \\&& +\Theta(a)+\Theta(\lambda-b)+\Theta(b)-2] 
      -\frac{\eta}{a-b}\frac{1}{2\pi}[\Theta(\lambda-b)  \nonumber \\ && +\Theta(b)
        -\Theta(\lambda-a)-\Theta(a)].
\end{eqnarray}
Evaluation of the radial integrals and addition of the four trace terms of Eq. (\ref{Eq:T_yy}) (see Appendix \ref{Appendix}) leads to,
\begin{eqnarray}\label{Eq.TraceComplete}
    T && (\e)=\frac{v_y}{v_x}\frac{\pi}{(2\pi)^2}\Biggl\{\int_0^{2\pi}\frac{\cos^2\theta}{1-\gamma^2\sin^2\theta}\Bigl(\frac{\w}{4}-\frac{\e}{2}\g\sin\theta\Bigr) \nonumber \\&&\times\delta_\eta \Bigl(\e-\frac{\w}{2}\g\sin\theta\Bigr)\frac{d\theta}{\pi}+\frac{\eta}{\pi\w}\varphi_{yy}^{tra}(\g)\Biggr\} \Theta\Bigl(\la-\frac{\w}{2}\Bigr).
\end{eqnarray}

The first term in $T(\e)$ is related to interband scattering, while the second term describes intraband scattering. An overall scaling factor of $v_y/v_x$ is introduced due to the anisotropy, and the function $\varphi_{yy}^{tra}(\g)$ enhances the intraband term as a consequence of the tilt in the energy dispersion. It is given by,

\begin{equation}\label{Eq.PhiIntra}
    \varphi^{tra}_{yy}(\g)=\frac{2}{\g^2}[(1-\g^2)(\sqrt{1-\g^2}-1)+\g^2].
\end{equation}

\begin{figure}[b]
\includegraphics[width=0.4\textwidth]{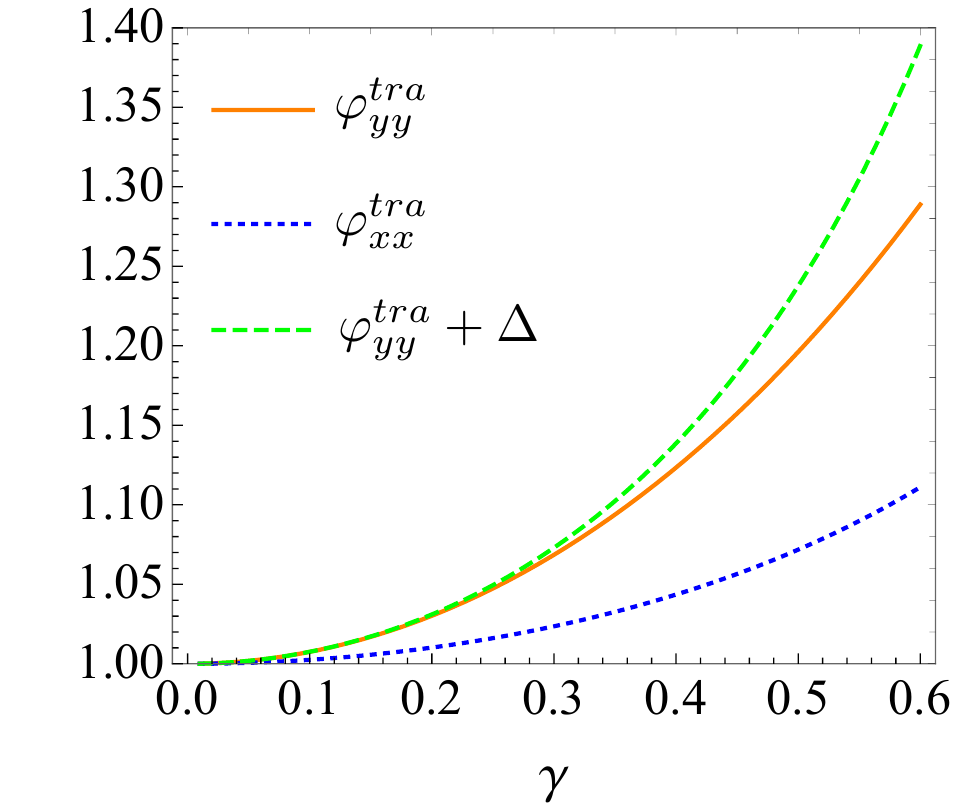}
\caption{\label{Fig.TiltFactors} A comparison of the different adimensional factors that appear in the conductivity due to the tilt strength $\g\neq 0$ of the energy dispersion. When a tilt is introduced in the $y$ direction, $\varphi^{tra}_{yy}(\g)$ enters the intraband contribution of $\sigma_{yy}$, while $\varphi^{tra}_{xx}(\g)$ enters the intraband contribution of $\sigma_{xx}$. The factor $\varphi_{yy}^{tra}+\Delta$ enters the minimal conductivity of Eq. (\ref{Eq.MinSigmaxxLim2}). All of these factors reduce to unity for a dispersion with no tilt. }
\end{figure}

We plot $\varphi^{tra}_{yy}(\gamma)$ in Fig. \ref{Fig.TiltFactors}. The magnitude of the intraband scattering term increases with the tilt; it reduces to that of graphene for $\g=0$. For the case of Pmmn-8 borophene  \cite{Zabolotskiy}, $\g=0.46$, resulting in  $ \varphi^{tra}_{yy}(\g)\approx 1.16$.

We can now calculate the temperature dependent conductivity by substituting Eq. (\ref{Eq.TraceComplete}) into Eq. (\ref{Eq:sigmadef}) (see Appendix \ref{Ap. Final sigma yy}).

Finally, we obtain that for $\w<2\lambda$,
\begin{eqnarray}\label{Eq.Sigmayy}
    \sigma_{yy}&& \sim  \frac{v_y}{v_x}\Biggl\{\varphi_{yy}^{ter}(\g,\beta\w)\frac{\pi e^2}{8h}\tanh\Bigg(\frac{\beta\w}{4}\Bigg) \nonumber \\&&
 +\varphi_{yy}^{tra}(\g)\frac{e^2}{h}\frac{\beta\eta}{(\beta \w)^2}\log\Biggl[\frac{1+\tanh^2(\beta\w/4)}{1-\tanh^2(\beta\w/4)}\Biggr]\Biggr\}
\end{eqnarray}
 and the conductivity vanishes for $\w>2\lambda$. Notice that in the previous equation we defined the interband scattering factor  as, 
\begin{eqnarray}\label{Eq.PhiInterDef}
    && \tanh \Bigg(\frac{\beta\w}{4}\Bigg)\varphi_{yy}^{ter}(\g,\beta\w) = \nonumber\\ &&\int_0^{2\pi}\cos^2\theta   \frac{\sinh(\beta\w/2)}{\cosh(\beta\w/2)+\cosh(\beta\frac{\w}{2}\g\sin\theta)}\frac{d\theta}{\pi}
\end{eqnarray}

After taking an expansion in 
$\gamma$, we obtain
\begin{equation}\label{Eq.PhiInterApprox}
    \varphi_{yy}^{ter}(\g,\beta\w)=1-\frac{\g^2}{4}\left(\frac{\beta\w}{4}\right)^2\text{Sech}^2\Bigg(\frac{\beta\w}{4}\Bigg) + O(\g)^4
\end{equation}

For $\g=0$ we recover the case of graphene, as $\varphi_{yy}^{ter}=1$. However, unlike in the intraband scattering factor, $\varphi_{yy}^{ter}$ is not only a function of $\g$, but of $\beta\w$ as well. 

The tilting has no effect in the interband conductivity in the limits $\beta\w \rightarrow 0$ and $\beta\w \rightarrow \infty$, as in both cases $\varphi_{yy}^{ter}(\g,\beta\w)=1$, just as in the case of no tilt ($\g=0$). For finite values of $\beta\w$, $\varphi_{yy}^{ter}$ decreases with $\g$
in contrast to $\varphi_{yy}^{tra}(\g)$, which monotonically increases.

\begin{figure}
    \centering
    \begin{minipage}{0.45\textwidth}
        \centering
        \includegraphics[width=0.75\textwidth]{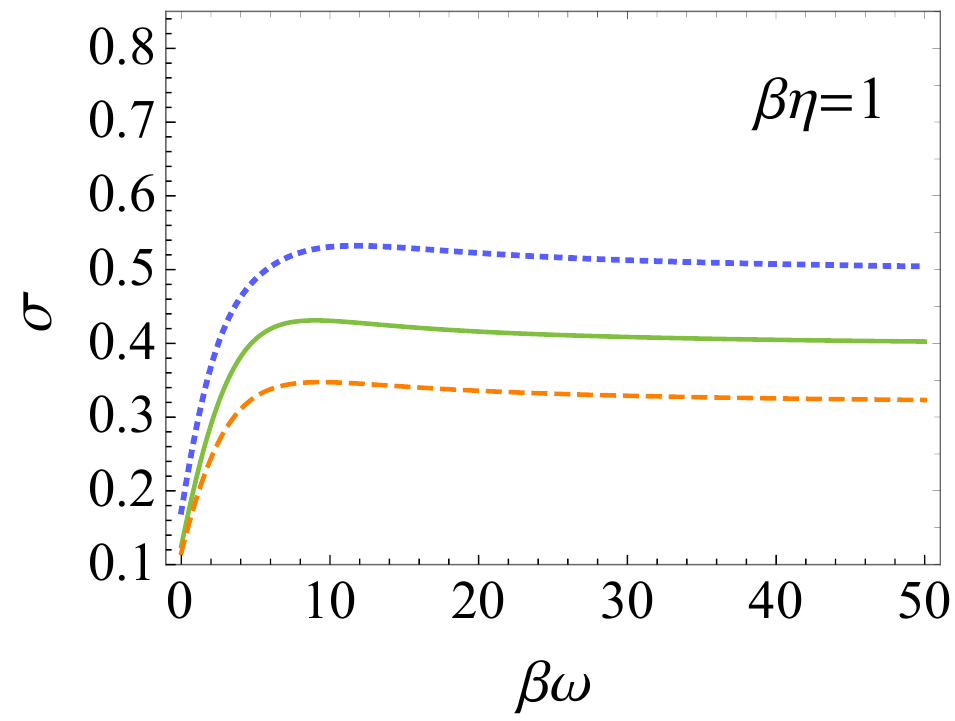} 
        \subcaption{}
    \end{minipage}\hfill
    \begin{minipage}{0.45\textwidth}
        \centering
        \includegraphics[width=0.75\textwidth]{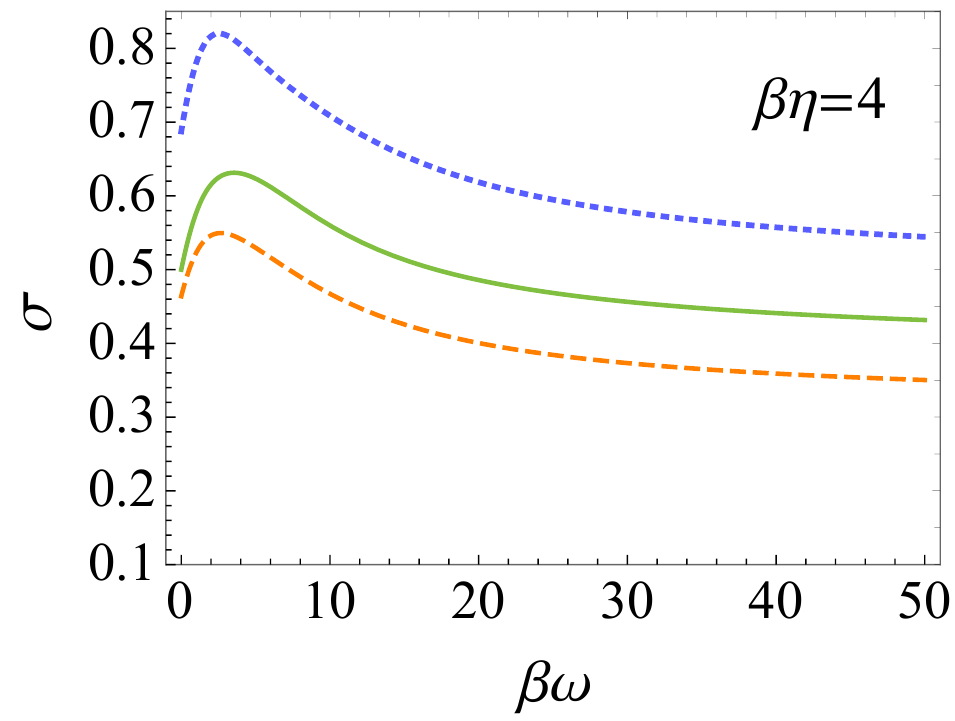} 
        \subcaption{}
    \end{minipage}
    \caption{ \label{Fig_Tot_ConductivityYY} Conductivity of 8-Pmmn borophene in the tilt axis $\sigma_{yy}$ (dashed curves), and the perpendicular axis $\sigma_{xx}$ (dotted curves), compared to that of graphene (solid curves) for the constant rates (a) $\beta\eta=1$ and (b) $\beta\eta=4$  assumed constant. Notice the anisotropy with respect to graphene. In the low-frequency limit for a fixed $T$, the conductivity depends upon the scattering. In the high-frequency region, it reaches the same limit for different amounts of scattering as happens with graphene. The numerical values for $v_x$, $v_y$ and $v_t$ were taken from \cite{Zabolotskiy}.}
\end{figure}

\begin{figure}[b]
\includegraphics[width=0.45\textwidth]{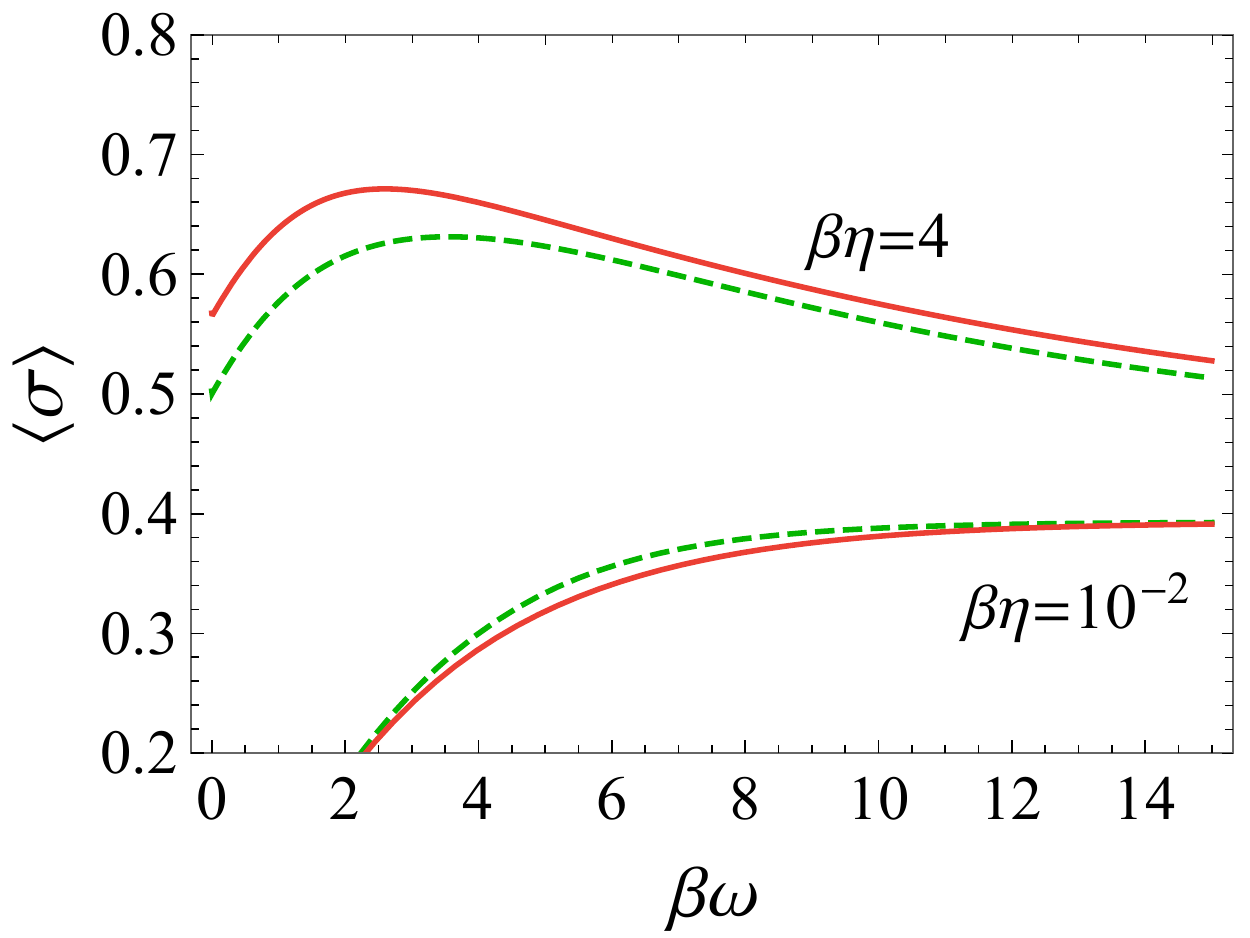} 
\caption{\label{Fig:ConductivityIso} Comparison of 
  graphene's conductivity (dashed lines) vs the geometric mean $\langle\sigma\rangle=\sqrt{\sigma_{xx}\sigma_{yy}}$ of 8-Pmmn borophene (solid lines) at different scattering rates $\beta\eta=10^{-2}$ (lower curves) and $\beta\eta=4$ (upper curves).}
\end{figure}

In Fig. \ref{Fig_Tot_ConductivityYY} we present the resulting $\sigma_{yy}$  (and also $\sigma_{xx}$, see next section for details on the calculation) as a function of $\beta \omega$ for the case of 8-Pmmn borophene using two different values for the scattering, $\beta \eta=1$ and  $\beta \eta=4$. For comparison proposes, we also plot graphene's conductivity. The predicted conductivity for 8-Pmmn borophene is smaller than that of graphene in the tilt direction, and larger in the perpendicular direction. In this case, the scaling factors $v_{\mu}/v_{\nu}$ dominate over the tilt factors $\varphi_{\mu\mu}^{intra}$ and $\varphi_{\mu\mu}^{inter}$. In order to show the effect introduced purely by the tilt, in Fig. \ref{Fig:ConductivityIso} is shown a comparison between graphene's conductivity and the geometric average $\langle\sigma\rangle=\sqrt{\sigma_{xx}\sigma_{yy}}$ for borophene, which is independent of $v_x$ and $v_y$. We can see that for the high-frequency limit, the mean geometrical conductivity is the same as in graphene.

As in graphene, samples under typical experimental situations have an appreciable spontaneous doping which is able to reduce the transition strength due to state blocking  \cite{Mak}. This can be accounted for by introducing a nonzero chemical potential $\mu$ relative to the Dirac point, which essentially shifts the peak of $\sigma$, giving a vanishing conductivity for values of $\w<2\mu$ and having practically no effect when $\w>2\mu$ \cite{Mak,Verma}.

Notice that apart from $\varphi_{yy}^{ter}$ and $\varphi_{yy}^{tra}$, the anisotropy in the energy dispersion introduces an additional scaling factor $v_y/v_x$ to $\sigma_{yy}$, in agreement with calculations using the Landauer formalism \cite{Trescher} or in a semi-Dirac material \cite{Carbotte}.

\subsection{\label{Subsec. Sigmaxx} Conductivity in the direction perpendicular to the tilt}

The calculation of $\sigma_{xx}$ can be performed by following the same steps as those presented in the previous section. The result is,

\begin{eqnarray}\label{Eq.Sigmaxx}
    \sigma_{xx} && \sim \frac{v_x}{v_y}\Biggl\{\varphi_{xx}^{ter}(\g,\beta\w)\frac{\pi e^2}{8h}\tanh\Biggl(\frac{\beta\w}{4}\Biggr) \nonumber\\ &&
 +\varphi_{xx}^{tra}(\g)\frac{e^2}{h}\frac{\beta\eta}{(\beta \w)^2}\log\Biggl[\frac{1+\tanh^2(\beta\w/4)}{1-\tanh^2(\beta\w/4)}\Biggr]\Biggr\}
\end{eqnarray}

The intraband scattering factor is, 
\begin{equation}\label{Eq.ChiIntraDef}
    \varphi_{xx}^{tra}(\g)=\int_0^{2\pi}\frac{1}{2\pi}\bigg\{\frac{\cos^2\theta}{\g_+}+\frac{\cos^2\theta}{\g_-} \bigg\}d\theta=\frac{2}{\g^2}[1-\sqrt{1-\g^2}]
\end{equation}
This function is analogous to $\varphi_{yy}^{tra}$; it also grows from unity as $\g$ increases, but it takes lower values. Both  $\varphi_{xx}^{tra}$ and $\varphi_{yy}^{tra}$ are compared in Fig. \ref{Fig.TiltFactors}. 

The interband scattering factor is given by,
\begin{equation}\label{Eq.ChiInterApprox}
    \varphi_{xx}^{ter}(\g,\beta\w)=1-\frac{3\g^2}{4}\left(\frac{\beta\w}{4}\right)^2\text{Sech}^2\Bigg(\frac{\beta\w}{4}\Bigg) + O(\g)^4
\end{equation}
It is worthwhile to remark that our Eqs. (\ref{Eq.Sigmayy}) and (\ref{Eq.Sigmaxx})  generalize Eq. (13) of Ref. \cite{Ziegler}, to which our expressions reduce for $v_y=v_x$ and $\g=0$. 

\section{\label{Sec.MinimalConductivity}Minimal conductivity}

In graphene, it is known that the minimal conductivity approaches different limits depending upon the scattering, frequency and temperature \cite{Mak,Ziegler}. In this section, we explore such limits. 
\subsection{\label{Subsec.Zero temperature}Zero temperature}
We put Eq. (\ref{Eq.TraceComplete}) into the conductivity of Eq. (\ref{Eq.sigmadefzeroT})  and obtain 
\begin{equation}\label{Eq.MinSigmayyLim1}
    \sigma_{yy}= \frac{v_y}{v_x}\frac{\pi}{8}\Bigg(1+ \frac{4\eta}{\pi\w}\varphi_{yy}^{tra}(\g)\Bigg)\frac{e^2}{h},
\end{equation}

Evaluating the limits discussed by Ziegler in Ref. \cite{Ziegler} at temperature zero we obtain, 

\begin{equation}\label{Eq.MinSigmayyLim2}
    \sigma_{yy,1}^{min}=\frac{v_y}{v_x}\varphi_{xx}^{tra}(\g) \frac{1}{\pi}\frac{e^2}{h},
\end{equation}
taking $\w \rightarrow 0$ in Eq. (\ref{Eq:T_yy}) then $\eta \rightarrow 0$,
\begin{equation}\label{Eq.MinSigmayyLim3}
    \sigma_{yy,2}^{min}=\frac{v_y}{v_x}\frac{\pi}{8}\frac{e^2}{h}\qquad \text{for $\eta\approx 0$},
\end{equation}
\begin{equation}\label{Eq.MinSigmayyLim4}
    \sigma_{yy,3}^{min}=\frac{v_y}{v_x}\frac{\pi}{4}\bigg(\frac{1+\varphi_{yy}^{tra}(\g)}{2}\bigg)\frac{e^2}{h} \qquad \text{for $\eta \approx \w$},
\end{equation}
these are to be compared to eqs. (15-17) of \cite{Ziegler}. The intraband factor that appears in the expression for $\sigma_{xx}$ enters $\sigma_{yy,1}^{min}$ and, while the minimal conductivities in eqs. (\ref{Eq.MinSigmayyLim2}) and (\ref{Eq.MinSigmayyLim4}) increase with the tilt, the one in Eq. (\ref{Eq.MinSigmayyLim3}) is not affected by it.

Quite analogous, for the zero temperature $\sigma_{xx}$ we obtain

\begin{equation}\label{Eq.MinSigmaxxLim1}
    \sigma_{xx}= \frac{v_x}{v_y}\frac{\pi}{8}\Bigg(1+ \frac{4\eta}{\pi\w}\varphi_{xx}^{tra}(\g)\Bigg)\frac{e^2}{h},
\end{equation}

\begin{equation}\label{Eq.MinSigmaxxLim2}
    \sigma_{xx,1}^{min}=\frac{v_x}{v_y}[\varphi_{yy}^{tra}(\g)+ \Delta (\g)]\frac{1}{\pi}\frac{e^2}{h}
\end{equation}
taking $\w \rightarrow 0$ in Eq. (\ref{Eq:T_yy}) then $\eta \rightarrow 0$,
\begin{equation}\label{Eq.MinSigmaxxLim3}
    \sigma_{xx,2}^{min}=\frac{v_x}{v_y}\frac{\pi}{8}\frac{e^2}{h}\qquad \text{for $\eta\approx 0$},
\end{equation}
\begin{equation}\label{Eq.MinSigmaxxLim4}
    \sigma_{xx,3}^{min}=\frac{v_x}{v_y}\frac{\pi}{4}\bigg(\frac{1+\varphi_{xx}^{tra}(\g)}{2}\bigg)\frac{e^2}{h} \qquad \text{for $\eta \approx \w$},
\end{equation}
having introduced $\Delta(\g)$ such that $\varphi_{yy}^{tra}(\g)+\Delta(\g)=2[(1-\g^2)^{-1/2}-1]/\g^2$. These values are plotted and compared in Fig. \ref{Fig.TiltFactors}. We note that for small values of $\g$, $\Delta(\g)\approx 0$ and then all of eqs. (\ref{Eq.MinSigmaxxLim1}-\ref{Eq.MinSigmaxxLim4}) are symmetrical to eqs. (\ref{Eq.MinSigmayyLim1}-\ref{Eq.MinSigmayyLim4}) under the interchange $\varphi_{xx}^{tra} \leftrightarrow \varphi_{yy}^{tra}$ and $v_y/v_x \leftrightarrow v_x/v_y$.
\subsection{\label{Subsec. Freq and Temp in MinCoductivity}Frequency and temperature dependence}
Now, considering the asymptotic regimes $\beta\w\rightarrow0$ and $\beta\w\rightarrow\infty$
for the conductivity in Eq. (\ref{Eq.Sigmayy}) leads to 
\begin{equation}\label{Eq.SigmayyOfT}
    \sigma_{yy}^\prime \sim \frac{v_y}{v_x}\frac{e^2}{8h}\Bigg\{\begin{array}{cc}
        \text{$ \;\,\varphi_{yy}^{tra}(\g)\times\beta\eta$ \qquad \qquad for $\beta\w\sim 0$}, \\
         \text{$\pi +\varphi_{yy}^{tra}(\g)\times 4\beta\eta/\beta\w$ \quad for $\beta\w\sim\infty$}.
    \end{array}
\end{equation}

Similarly, for the conductivity in Eq. (\ref{Eq.Sigmaxx}),
\begin{equation}\label{Eq.SigmaxxOfT}
    \sigma_{xx}^\prime \sim \frac{v_x}{v_y}\frac{e^2}{8h}\Bigg\{\begin{array}{cc}
        \text{$ \;\,\varphi_{xx}^{tra}(\g)\times\beta\eta$ \qquad \qquad for $\beta\w\sim 0$}, \\
         \text{$\pi +\varphi_{xx}^{tra}(\g)\times 4\beta\eta/\beta\w$ \quad for $\beta\w\sim\infty$}.
    \end{array}
\end{equation}

\section{\label{Sec.Discussion}Discussion of the results}
We first discuss our results in the zero temperature, dc limit.
The minimal conductivities in Eqs. (\ref{Eq.MinSigmayyLim2}) and (\ref{Eq.MinSigmaxxLim2}) can be written as

\begin{equation}
    \sigma_{yy,1}^{min}=\frac{v_y}{v_x}\frac{1}{\pi}\frac{e^2}{h}\frac{2}{\g^2}\left(\frac{1}{\sqrt{1-\g^2}}-1\right)\sqrt{1-\g^2}
\end{equation}

\begin{equation}
    \sigma_{xx,1}^{min}=\frac{v_x}{v_y}\frac{1}{\pi}\frac{e^2}{h}\frac{2}{\g^2}\left(\frac{1}{\sqrt{1-\g^2}}-1\right).
\end{equation}
We notice that both $\sigma_{xx,1}^{min}$ and $\sigma_{yy,1}^{min}$ increase with the tilt parameter $\g$, while $\sigma_{xx,1}^{min}>\sigma
_{yy,1}^{min}$ and only $\sigma_{xx,1}^{min}$ diverges as $\g\rightarrow 1$ \cite{Suzumura,Goerbig2019}. This limit coincides with recent calculations of the static conductivity using the covariant Boltzmann equation \cite{Goerbig2019}. The fact that both $\sigma_{xx,1}^{min}$ and $\sigma_{yy,1}^{min}$ grow with the tilt of the dispersion can be attributed to the increase in the density of states with the tilt parameter $\g$,
\begin{equation}\label{Eq.DOSintra}
    \rho(E)=\sum_{\boldsymbol{k}}^{|k|<k_F} \delta(E_\pm(\boldsymbol{k})-E)=\frac{|E|}{2\pi v_x v_y}(1-\g^2)^{-3/2}
\end{equation}
To account for the anisotropy we point out that the constant-energy cross sections of the dispersion cone describe ellipses in momentum space given by $(v_x k_x)^2/A^2+(v_y k_y -h)^2/B^2=1$ with $h=\pm \g E/(1-\g^2)$, $A=E/\sqrt{1-\g^2}$ and $B=E/(1-\g^2)$. As the tilt increases, the eccentricity of the isoenergetic ellipses becomes more pronounced and scattering events across the shorter axis (here the x-axis) become more probable \cite{Goerbig2019,Suzumura} than along the tilt axis (here the y-axis). 
We further note that in the purely tilted case ($v_x=v_y$)  the ratio between the lengths of the ellipse's semi-axes $B/A$ is precisely equal to the ratio between these conductivities
\begin{equation}
    \frac{\sigma_{xx,1}^{min}}{\sigma_{yy,1}^{min}}=\frac{1}{\sqrt{1-\g^2}}=\frac{B}{A}.
\end{equation}
When anisotropy ($v_y\neq v_x$) is introduced, we obtain a direction-dependent scaling of the components $\sigma_{\nu\nu}$ of the form $v_\nu/v_\mu$ which is in agreement with calculations using the Landauer formalism \cite{Trescher} or in a semi-Dirac material \cite{Carbotte}.

\begin{figure}[b]
\includegraphics[width=0.45\textwidth]{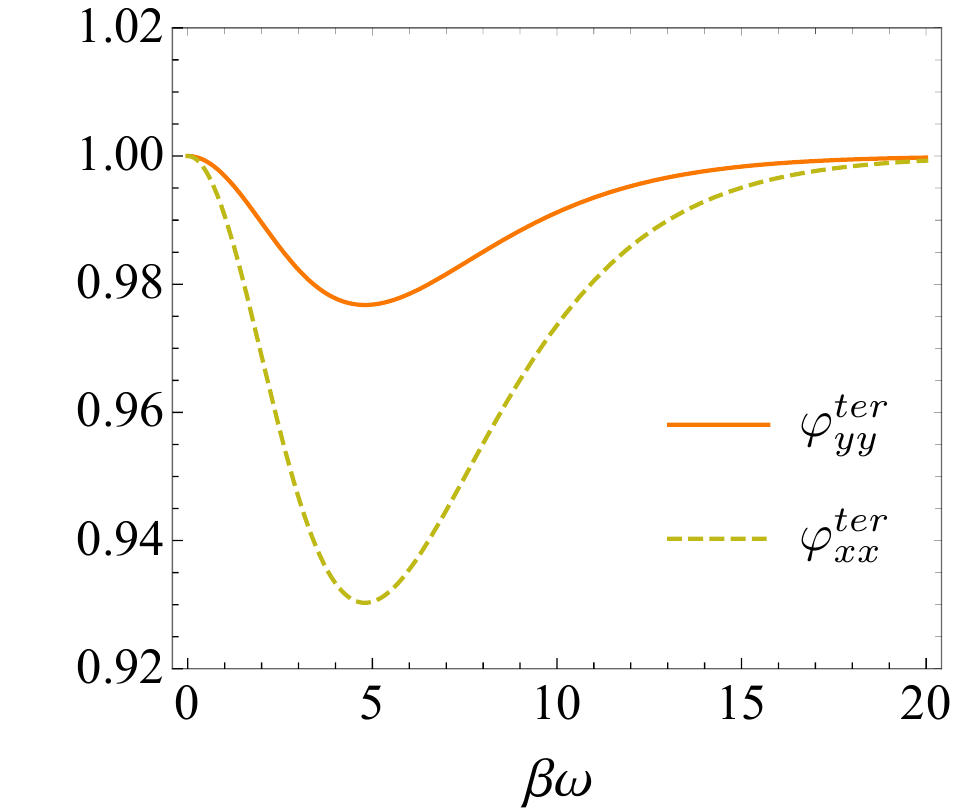}
\caption{\label{Fig:PhiInter} Comparison of the adimensional factors that appear in the interband components of the conductivity due to a tilt strength $\g=0.46$ which has been reported for 8-Pmmn borophene's energy dispersion \cite{Zabolotskiy}. $\varphi_{xx}^{ter}$ enters the interband component of $\sigma_{xx}$ and $\varphi_{yy}^{ter}$ enters the interband component of $\sigma_{yy}$. A minimum is observed for $\beta \omega\approx4$ for any $\gamma \neq 0$.  }
\end{figure}

\begin{figure*}
\includegraphics[width=0.8\textwidth]{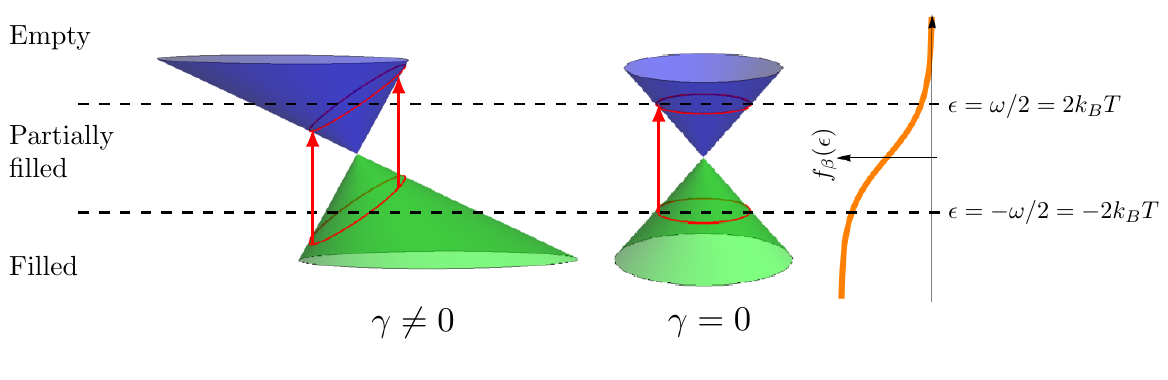}

\caption{Comparison of interband transitions occurring in a material with a tilted ($\gamma\neq 0$) vs a non-tilted ($\gamma=0$) dispersion cone when $\beta\omega\approx 4$. The red contours in the cones indicate the states the can participate in direct interband transitions i.e., states that satisfy $E_+(\boldsymbol{k})-E_-(\boldsymbol{k})
=\omega$. When $\gamma\neq 0$, the contours tilt and half of their perimeter lays in the partially-filled energy region, where the total number of states available for transitions decreases. This reduces the interband conductivity around $\beta\omega\approx 4.$
 \label{DispersionAndMu}}
\end{figure*}

Our results show that in the zero temperature limit interband transitions are not affected by the tilt in the dispersion. As shown in Fig. \ref{DispersionAndMu}, such results comes out from the global tilting of the transitions. This can also be readily verified by using the Fermi golden rule, i.e., we introduce the electric field $\boldsymbol{E}$ as a perturbation $\delta H$ to the Hamiltonian in Eq. (\ref{Eq:Hamiltonian}) as 
\begin{equation}
    H=\boldsymbol{\sigma}^\prime\cdot(\boldsymbol{p}-\frac{e}{c}\boldsymbol{A})\equiv H_0+\delta H
\end{equation}
where $\boldsymbol{\sigma}^\prime=(v_x\sigma_x,v_y(\sigma_y+\gamma\sigma_0))/\hbar$, $\boldsymbol{p}=\hbar \boldsymbol{k}$ and $\partial \boldsymbol{A}/\partial t = -c\boldsymbol{E}$. The perturbation is then given by \cite{KATSNELSON200720}
\begin{equation}\label{Eq.Perturbation}
    \delta H=\frac{ie}{2\omega}\boldsymbol{\sigma}^\prime\cdot\boldsymbol{E}
\end{equation}
According to the Fermi golden rule, the absorption energy per unit time for direct interband transitions $E_-(\boldsymbol{k})\rightarrow E_+(\boldsymbol{k})$ between states with an energy difference $\Delta E=\omega$ is \cite{KATSNELSON200720,Sakurai}
\begin{equation}\label{Eq.FermiGoldenRule}
    W= g_d\sum_{\boldsymbol{k}}4 \pi\omega|\langle\Psi_+(\boldsymbol{k})|\delta H|\Psi_-(\boldsymbol{k})\rangle|^2\delta(E_+(\boldsymbol{k})-E_-(\boldsymbol{k})-\omega)
\end{equation}

where $g_d=4$ takes into account valley and spin degeneracy. It can be easily seen that the transition amplitude $\langle\Psi_+(\boldsymbol{k})|\delta H|\Psi_-(\boldsymbol{k})\rangle$ is independent of $\gamma$, as the term proportional to $\sigma_0$ is eliminated due to the orthogonal character of the basis. Moreover, we make the observation that the density of states implied in Eq. (\ref{Eq.FermiGoldenRule}) is not the same as the one in Eq. (\ref{Eq.DOSintra}). This is due to the tilt, as the states that participate in interband transitions between states with a given energy difference $\Delta E= \omega$ do not lay on an isoenergetic curve in the dispersion (Fig. \ref{Fig:DispersionCone}), unlike in graphene ($\g=0$), where the density of states that enters the expression for the transition rate is simply $\rho(\omega/2)$. Rather, the density of final states for direct interband transitions with $\Delta E 
=\omega$ is given by (see Appendix \ref{Ap. Absorption coef})
\begin{equation}\label{Eq.DOSinterband}
    D(\omega/2)\equiv 
2 g_d \sum_{\boldsymbol{k}}\delta(E_+(\boldsymbol{k})-E_-(\boldsymbol{k})-\omega)=\frac{\omega}{\pi v_x v_y }
\end{equation}
which is independent of $\g$. In the isotropic case ($v_x=v_y$), $D(\omega/2)$ is exactly graphene's density of final states  $\rho(\omega/2)$ for interband transitions with $\Delta E=\omega$.

The final expressions for the absorption coefficient on each direction are
\begin{equation}\label{Eq. Absorption coef}
    \frac{W_{x}}{W_i}=\frac{v_x}{v_y}\frac{\pi e^2}{\hbar c}, \quad \frac{W_{y}}{W_i}=\frac{v_y}{v_x}\frac{\pi e^2}{\hbar c}
\end{equation}

where $W_i=c|\boldsymbol{E}|/4\pi \hbar$ is the incident energy flux \cite{KATSNELSON200720}. When $v_x=v_y$, this direction-dependent optical absorption coefficient reduces to the isotropic constant value of $\pi e^2 /\hbar c = \pi \alpha$ measured in graphene  \cite{Mak,Nair}. This  shows that in the $T\rightarrow 0$ limit, the tilt in the dispersion has no effect on interband transitions, a fact consistent with the expressions for the factors $\varphi_{\mu\mu}^{ter}$ obtained from the Kubo formula (Fig. \ref{Fig:PhiInter}).
Notice that for intraband transitions, the relevant density of states is that in Eq. (\ref{Eq.DOSintra}), while for analyzing the interband transitions, the $\gamma$-independent density of states in Eq. (\ref{Eq.DOSinterband}) should be used. The former is related to the states in a isoenergetic cross-section of the dispersion (which is to be taken into account in elastic scattering events) and the latter is related to the states that can participate in direct interband transitions.

Thus, our results show that the intraband  contribution to the conductivity, which is related to the Drude peak, is enhanced by the tilt in the energy dispersion, and this effect is highly anisotropical.

On the other hand, for the interband conductivity we obtain that both components $\sigma_{yy}$ and $\sigma_{xx}$ decrease in the limit where $\beta\omega$ takes a finte value. Moreover, if we express Eq. (\ref{Eq.ChiInterApprox}) as $\varphi_{xx}^{ter}=1-\delta\varphi_{xx}^{ter}$, and do similarly for $\varphi_{yy}^{ter}$, comparing this equation with Eq. (\ref{Eq.PhiInterApprox}), we note that,
\begin{equation}\label{ChiVsPhi}
     \delta\varphi_{xx}^{ter}+\delta\varphi_{yy}^{ter}=\g^2c_v/k_B
\end{equation}
up to $O(\g)^4$ terms, where $c_v$ is the heat capacity (per particle) of a two level system with an energy gap of $\Delta E=\w/2$,
\begin{equation}\label{Eq.HeatCapacity}
    c_v=k_B\left(\frac{\beta \Delta E}{2}\right)^2\text{Sech}^2\left(\frac{\beta \Delta E}{2}\right)
\end{equation}

In Fig. \ref{Fig:PhiInter}, we present the resulting curves for $\varphi_{xx}^{ter}$ and $\varphi_{yy}^{ter}$. Notice that as happens with the specific heat of a two-level system, there is a maximum for $c_v$ which in  this case results in a minimum for $\varphi_{xx}^{ter}$ and $\varphi_{yy}^{ter}$ as a function of $\beta \omega$. This occurs at $\beta \omega \approx 4$  as obtained from the derivative of Eq. (\ref{Eq.HeatCapacity}).  A sketch of its explanation is presented in Fig. \ref{DispersionAndMu}. The minimum arises as an interplay between the partially-filled energy region in the Fermi distribution, which extends approximately from $-2 k_{B}T$ to $+2 k_{B}T$ around the chemical potential, with the tilted cross-sections in the cone that carry interband transitions for a given $\omega$ . In the case of graphene ($\gamma=0$), the interband conductivity takes a constant value for $\beta\omega \gtrsim 4$ and it starts to decrease when $\beta\omega \lesssim 4$ \cite{Ziegler}, as the contours in the dispersion cone that participate in interband transitions start to enter the partially-filled energy region, where the number of total states available for transitions is reduced. When $\gamma \neq 0$, as in 8-Pmmn borophene, these contours are tilted and as a result, when $\beta\omega\approx 4$, already half of their perimeter is in the region of partially filled states. This accounts for the reduction of the mean geometrical conductivity of 8-Pmmn borophene with respect to that of graphene in a wide range around $\beta\omega\approx4$ shown in Fig. \ref{Fig:ConductivityIso} for $\beta\eta=10^{-2}$.

\subsection{\label{Subsec. Conclusions} Conclusions}
We have investigated the temperature-dependent optical conductivity of anisotropic tilted Dirac semimetals using the Kubo formula and discussed our results in the context of 8-Pmmn borophene. The effects of the tilting on interband and intraband scattering were analyzed in detail. We found direction-dependent scaling factors that appear due to the anisotropy of the energy dispersion and an anisotropic increase of the intraband conductivity with the tilt strength $\g$, which can be attributed to the deformation of the constant-energy contours of the dispersion cone into ellipses. In the zero-temperature, dc limit, our results are in agreement with recent calculations of the static conductivity using the covariant Boltzmann equation \cite{Goerbig2019}. We also found most of the limits leading to minimal conductivities that increase with the tilt strength. Our results reproduce those of graphene reported in \cite{Ziegler} for the particular case of an isotropic energy dispersion with no tilt.
Moreover, the conductivity is similar to that found in graphene but weighted by the anisotropy in such a way that the mean geometrical conductivity  is  the  same  as  in  graphene  for  the  high  frequency or low temperature limit. This is a consequence of the fact that in such limit, interband transitions are not affected by the tilt in the dispersion, a result that was verified by a direct use of the Fermi golden rule. Finally, as the temperature was raised, a minimum of the interband dispersion was observed as a result of the interplay between tilting and the derivative of the Fermi distribution widening with temperature. Such minimum can be tested by a suitable optical experiment.

\begin{acknowledgments}
This work was supported by DGAPA project IN102717. S. A. Herrera was supported by a CONACyT MSc. scholarship. 
\end{acknowledgments}

\appendix

\section{\label{Appendix} Calculation of the trace factor $T(\epsilon)$ in the expression for $\sigma_{yy}$}

In this section we calculate the trace factor in Eq. (\ref{Eq.sigmadefzeroT}). For $\nu=y$ we have defined
\begin{eqnarray}
T(\epsilon)&&=-\text{Tr}\{[H,r_y]\delta(H-\omega/2-\epsilon)[H,r_y]\delta(H+\omega/2-\epsilon)\} \nonumber \\ 
&&=\int \text{Tr}_2 \biggl[\frac{\partial H}{\partial k_y}\delta \bigl(H-\frac{\omega}{2}-\epsilon\bigr) \frac{\partial H}{\partial k_y}\delta\bigl(H+\frac{\omega}{2}-\epsilon\bigr)\biggr]\frac{d^2k}{(2\pi)^2} \nonumber
\end{eqnarray}
Expanding the trace over the pseudospin degree of freedom we get 
 \begin{eqnarray}\label{Ap.TraceExpansion}
    T(\epsilon)&&=\int \{\bp \Lambda_1\kp \bp\Lambda_2\kp \nonumber \\&& +\bp\Lambda_1\km\bmm\Lambda_2\kp  +\bmm\Lambda_1\kp\bp\Lambda_2\km \nonumber \\ && +\bmm\Lambda_1\km\bmm\Lambda_2\km\}\frac{d^2k}{(2\pi)^2},
 \end{eqnarray}
 where we have defined the operators
$$\Lambda_1=\frac{\partial H}{\partial k_y}\delta(H-\frac{\omega}{2}-\epsilon), \;\;\;\Lambda_2=\frac{\partial H}{\partial k_y}\delta(H+\frac{\omega}{2}-\epsilon).$$

Substitution of the elements of the current operator in Eq. (\ref{Eq.CurrentOp}) into Eq. (\ref{Ap.TraceExpansion}) leads  to Eq. (\ref{Eq:T_yy}).
In the following, we will assume $-\omega/2\leq\epsilon\leq\omega/2$, which is justified for low temperatures.
\subsection{Expression for $T_{+}^{ter}(\epsilon)$}

From Eq. (\ref{Eq.T ter}),
\begin{equation}\label{Eq. T ter +}
    T_{+}^{ter}(\epsilon)=\int_0^{2\pi} l(\theta)\int_0^\lambda \delta_{\eta}\Big(\xi +\frac{\bar{\epsilon}_{+}}{\gamma_{-}}\Big)\delta_{\eta}\Big(\xi + \frac{\bar{\epsilon}_{-}}{\gamma_{+}}\Big)  \frac{ \xi d\xi d\theta}{(2\pi)^2}. \nonumber
\end{equation}
Comparing with Eq. (\ref{Eq.DoubleDeltaInt}) we identify $a=-\bar\epsilon_+/\gamma_-$ and $b=-\bar\epsilon_-/\gamma_+$. As $a,b\leq0$, we have $\Theta(a)=\Theta(b)=0$ and $\Theta(\lambda-a)=\Theta(\lambda-b)=1$. Therefore, according to Eq. (\ref{Eq.DoubleDeltaInt}), $T_{+}^{ter}(\epsilon)=0$.

\subsection{Expression for $T_{-}^{ter}(\epsilon)$}
From Eq. (\ref{Eq.T ter}),

\begin{equation}\label{Eq.T ter -}
    T_{-}^{ter}(\epsilon)=\int_0^{2\pi}\int_0^\lambda  l(\theta)\delta_{\eta}\Big(\xi -\frac{\bar{\epsilon}_{+}}{\gamma_{+}}\Big)\delta_{\eta}\Big(\xi - \frac{\bar{\epsilon}_{-}}{\gamma_{-}}\Big)\frac{ \xi d\xi d\theta}{(2\pi)^2} \nonumber
\end{equation}
Comparing to Eq. (\ref{Eq.DoubleDeltaInt}) we identify $a=\bar\epsilon_+/\gamma_+$ and $b=\bar\epsilon_-/\gamma_-$. As $a,b \geq 0$, we have $\Theta(a)=\Theta(b)=1$. Adding the inequalities $\lambda>a$ and $\lambda >b$ leads to $\Theta(\lambda-a)+\Theta(\lambda-b)=2\Theta(\lambda-\omega/2)$ and $\Theta(\lambda-a)-\Theta(\lambda-b)=0$. Therefore, according to Eq. (\ref{Eq.DoubleDeltaInt}),
\begin{eqnarray}\label{Eq.T ter -}
    T_{-}^{ter}(\epsilon)=\int_0^{2\pi}  l(\theta)
    \Bigl(\frac{\w}{4}-\frac{\e}{2}&&\g\sin\theta\Bigr) \delta_\eta \Bigl(\e-\frac{\w}{2}\g\sin\theta\Bigr)
     \nonumber \\ &&
     \times \frac{ d\theta}{(2\pi)^2}\Theta(\lambda-\omega/2).
\end{eqnarray}
\subsection{Expressions for $T_{+}^{tra}(\epsilon)$ and $T_{-}^{tra}(\epsilon)$ }
From Eq.(\ref{Eq.T tra}),

\begin{equation}\label{Eq.T tra +}
    T_{+}^{tra}(\epsilon)=\int_0^{2\pi}\int_0^\lambda
    g_{+}(\theta)\delta_{\eta} \Big(\xi -\frac{\bar{\epsilon}_{+}}{\gamma_{+}}\Big)\delta_{\eta} \Big(\xi + \frac{\bar{\epsilon}_{-}}{\gamma_{+}}\Big)\frac{ \xi d\xi d\theta}{(2\pi)^2}. \nonumber
\end{equation}
Comparing to Eq. (\ref{Eq.DoubleDeltaInt}) we identify $a=\bar\epsilon_+/\gamma_+$ and $b=-\bar\epsilon_-/\gamma_+$. As $a\geq0$ and $b\leq0$, we have $\Theta(a)=1$, $\Theta(b)=\Theta(\lambda-b)=0$.

On the other hand,
\begin{equation}\label{Eq.T tra -}
    T_{-}^{tra}(\epsilon)=\int_0^{2\pi}\int_0^\lambda
    g_{-}(\theta)\delta_{\eta} \Big(\xi +\frac{\bar{\epsilon}_{+}}{\gamma_{-}}\Big)\delta_{\eta} \Big(\xi - \frac{\bar{\epsilon}_{-}}{\gamma_{-}}\Big)\frac{ \xi d\xi d\theta}{(2\pi)^2}. \nonumber
\end{equation}

And in this case, comparing to Eq. (\ref{Eq.DoubleDeltaInt}) we identify $a=-\bar\epsilon_+/\gamma_-$ and $b=\bar\epsilon_-/\gamma_-$. As $a\leq0$ and $b\geq0$, we have $\Theta(a)=\Theta(\lambda-a)=0$, $\Theta(b)=1$.
Adding these two last expressions and solving the integral over $\theta$ leads to
\begin{eqnarray}\label{Eq. T tra sum}
T^{tra}&&\equiv T_{+}^{tra}(\epsilon)+T_{-}^{tra}(\epsilon) \nonumber \\ &&=\frac{\eta}{2\pi\omega}\int_0^{2\pi}[g_+(\theta)\gamma_+ + g_-(\theta)\gamma_-]\frac{d \theta}{(2\pi)^2}\times\Theta(\lambda-\omega/2) \nonumber \\
&&=\frac{v_y}{v_x}\frac{\eta/\omega}{(2\pi)^2}\times\varphi_{yy}^{ter}(\gamma)\times\Theta(\lambda-\omega/2)
\end{eqnarray}
where $\varphi_{yy}^{tra}(\gamma)$ is given in eq. (\ref{Eq.PhiIntra}).
Adding Eqs. (\ref{Eq. T ter +}) and (\ref{Eq. T tra sum}) leads to Eq. (\ref{Eq.TraceComplete}).

\section{\label{Ap. Final sigma yy}The final expression for $\sigma_{yy}$}

To obtain the final expression for the conductivity in Eq. (\ref{Eq.Sigmayy}), we substitute Eq. (\ref{Eq:T_yy}) into Eq. (\ref{Eq:sigmadef}) and write

\begin{equation}
    f_\beta \Bigl(\epsilon+\frac{\omega}{2}\Bigr)-f_\beta\Bigl(\epsilon-\frac{\omega}{2}\Bigr)=\frac{-\sinh(\beta\omega/2)}{\cosh(\beta\omega/2)+\cosh(\beta \epsilon)}. \nonumber
\end{equation}
Then the integral over $\epsilon$ is solved using 
 \begin{eqnarray}
        \frac{1}{\w}\int_{-\w/2}^{\w/2}\frac{\sinh(\beta\w/2)}{\cosh(\beta\w/2)+  \cosh(\beta\e) } &&d\e = \nonumber \\  \frac{4}{\beta\w}\text{arctanh}&&\Bigg[\tanh^2\Bigg(\frac{\beta\w}{4}\Bigg)\Bigg]\, \nonumber
\end{eqnarray}
together with $\text{arctanh}(x)=(1/2)\log[(1+x)/(1-x)]$. 
To solve the remaining integral over $\theta$ we make an expansion around $\gamma=0$ as shown in Eqs. (\ref{Eq.PhiInterDef}) and Eqs. (\ref{Eq.PhiInterApprox}).

\section{Optical absorption coefficient}\label{Ap. Absorption coef}
The total absorption energy per unit time shown in Eq. (\ref{Eq.FermiGoldenRule}) is usually expressed as  \cite{KATSNELSON200720}
\begin{equation}\label{Ap.Fermi energy}
W=\frac{2\pi}{\hbar}\overline{|\langle\Psi_+|\delta H|\Psi_-\rangle|^2}D(\omega/2)\hbar \omega
\end{equation}

where the average is taken over $2\pi$ and $D(\omega/2)$ is defined as the density of final states for transitions between states with an energy difference $\Delta E=\omega$,
\begin{equation}
    D(\omega/2)=2g_d\sum_{\boldsymbol{k}}\delta(E_+(\boldsymbol{k})-E_-(\boldsymbol{k})-\omega)
\end{equation}
The factor of $2$ has to be introduced because $D(\omega/2)=dN/d(\omega/2)=2\times dN/d\omega$. One can easily corroborate that this definition yields the familiar result of $D(\omega/2)=\omega/\pi v_F^2$ for graphene \cite{KATSNELSON200720} , where $E_\pm(\boldsymbol{k})=\pm v_F k$ (energy is in units of $\hbar$).
From Eq. (\ref{Eq.Perturbation}) we get
\begin{equation}
  \overline{|\langle\Psi_+|\delta H|\Psi_-\rangle|^2}=e^2 v_\nu^2 E_\nu^2/8\omega^2\hbar^2  
\end{equation}
and substituting into Eq. (\ref{Ap.Fermi energy}) leads to Eqs.(\ref{Eq. Absorption coef}).

\bibliography{apssamp}

\end{document}